\documentstyle[prl,aps]{revtex}

\newcommand{\nn}{\nonumber}
\newcommand{\eea}{\end{eqnarray}}
\newcommand{\bea}{\begin{eqnarray}}
\newcommand{\ba}{\begin{array}}
\newcommand{\ea}{\end{array}}
\newcommand{\be}{\begin{equation}}
\newcommand{\ee}{\end{equation}}
\newcommand{\bd}{\begin{description}} \newcommand{\ed}{\end{description}}
\newcommand{\bfig}{\begin{figure}} \newcommand{\efig}{\end{figure}}

\begin{document}
\draft
\preprint{HEP/123-qed}
\title{ Time to failure of hierarchical load-transfer models of fracture}
\author{M. V\'{a}zquez-Prada}
\address{Departamento de F\'{\i}sica Te\'{o}rica, Universidad de Zaragoza, 50009 Zaragoza,
Spain.}
\author{J. B. G\'{o}mez}
\address{Departamento de Ciencias de la Tierra,
 Universidad de Zaragoza, 50009 Zaragoza, Spain.}
\author{Y. Moreno\cite{byline}, and A. F. Pacheco} \address{Departamento de F\'{\i}sica
Te\'{o}rica, Universidad de Zaragoza, 50009 Zaragoza, Spain.}
 \date{\today}
 \maketitle
 \widetext
 \begin{abstract} The time to failure, $T$,  of dynamical models of fracture for
a hierarchical load-transfer geometry is  studied. Using a probabilistic strategy and 
juxtaposing hierarchical structures of height $n$, we devise an exact  method to compute 
$T$, for structures of height $n+1$. Bounding $T$, for large $n$, we are able to deduce 
that the time to failure tends to a non-zero value when $n$ tends to infinity. This 
numerical conclusion is deduced for both power law and exponential breakdown rules. 
\end{abstract}  


\section{Introduction}
\label{sec:level1}

Fracture in heterogeneous materials is a complex physical problem
for which a definite physical and theoretical treatment is still
lacking. By ``heterogeneous material''we understand a system whose
breaking properties (e.g., strengths, lifetimes) depend on time
and/or space in a random way \cite{hermann}. This randomness
arises from the many-body interactions among the constituent parts
of the system, each one having mechanical properties that can be
considered independent of --or at least weakly correlated with--
the properties of neighboring parts. The term disordered systems
is also used as a collective name for this kind of material. The
presence of disorder alters radically the way the rupture process
evolves compared to the single-crack growth mechanism operating in
homogeneous materials (such as glass or alloys). In heterogeneous
material (composites, ceramics, rocks, concrete) the process of
rupture begins with delocalized damage affecting the bulk of the
material, and consisting of an enormous number of microcracks
nucleated at random inside the system. This population of
microcracks evolves with time by coalescence and growth of
individual microcracks until the final rupture point of the system
is reached. In the very final stages, the process of coalescence
gives rise to a single (or a few) dominant crack(s) responsible
for the macroscopic failure of the material.

The analytical or even complete numerical solution of this complex
problem is prohibitive. Nevertheless, our understanding of
fracture in heterogeneous material has improved recently with the
development of simple algorithms to simulate the breaking process.
Most of these algorithms are based on percolation theory
\cite{stauffer} and include models of random resistor networks
\cite{arcan}, spring networks \cite{feng}, and beam networks
\cite{roux}. The standard way of solving these models is through a
more or less dense sampling of failure space via Monte Carlo
simulation. But fracture in heterogeneous materials is, from a
statistical viewpoint, a process critically dependent on the tails
of the failure distribution and these tails are naturally
difficult to sample using conventional Monte Carlo methods. It is
thus very important to develop a set of simple models which can be
analyzed either analytically or numerically, with precision and
with clear asymptotic behavior, in order to guide our
understanding of more complex models.

The load-transfer models belong to this group of simple,
stochastic fracture models amenable to either close analytical or
fast numerical solution, and whose output, spanning many orders of
magnitude in sample size, allows a precise characterization of the
asymptotic behavior. The collective name given to this type of
model is fiber-bundle models (FBM), because they arose in close
connection with the strength of bundles of textile fibers
\cite{daniels}, \cite{coleman}. Since Daniels' and Coleman's
seminal works, there has been a long tradition in the use of these
simple models to analyze failure of heterogeneous materials.

FBM come in two ``flavors'', static and dynamic. The static versions of FBM simulate the 
failure of materials by quasistatic loading, i.e., by a steady increase in the load over 
the system up to its macroscopic failure. One of the basic outputs is precisely the value 
of this ultimate strength. Time plays no role in these models, load $\sigma$ is the 
independent variable, and the strength of each element is considered to be an independent 
identically distributed random variable. On the other hand the dynamic FBM simulate 
failure by stress-rupture, creep-rupture, static-fatigue or delayed-rupture, i.e., a 
(usually) constant load is imposed over the system and the elements break by fatigue 
after a period of time. The time elapsed until the system collapses is the lifetime or 
time to failure of the set. Time acts as the independent variable, and the lifetime of 
each element is an independent identically distributed random quantity. 

There are three basic ingredients common to all FBM: first, a
discrete set of $N$ elements located on the sites of a
$d$-dimensional lattice; second, a probability distribution for
the failure of individual elements; and, third, a load-transfer
rule which determines how the load carried by a failed element is
to be distributed among the surviving elements in the set.

 The most common probability distribution function (second ingredient) used to
express the breaking properties of individual elements is
the Weibull distribution \cite{weibull}. For the static cases,
where the load , $\sigma$, is the independent variable, failure
statistics are described by the function $P(\sigma)= 1-exp \lbrace
-(\sigma/\sigma_0)^\rho \rbrace$. Here, $\sigma_0$ is a reference
strength and $\rho$ is the so-called Weibull index or shape
parameter, which in essence controls the variance in the strength
thresholds. For the dynamic cases, with time as the independent
variable, things are more complicated because the failure of each
element is sensitive to both the elapsed time {\it and} its load
history. The probability $P(t;\sigma(t))$ of a single element
failing at time $t$ after suffering the load history $\sigma(t)$
is of the form \cite{coleman}
\begin{equation} \label{eq:1.1}
P(t;\sigma(t))=1-exp \left\{ - \int_0^{t}\kappa_j \left[
\sigma(\tau) \right] d\tau  \right\} ,
\end{equation}
where $\kappa_j(x)$, $j=1,2$, is the hazard rate or breaking rule.
To impart to Eq.~(\ref{eq:1.1}) the commonly observed Weibull
behavior of real materials under constant load a {\it power-law
breaking rule} is used \cite{coleman2}:
\begin{equation}  \label{eq:1.2}
\kappa_1(\sigma)=\nu_0 \left( \frac{\sigma}{\sigma_0}
\right)^\rho.
\end{equation}
Here, $\nu_0$ is the hazard rate (number of casualties per unit of time) under the unit
load $\sigma_0$. For constant load, inserting Eq.~(\ref{eq:1.2}) into Eq.~(\ref{eq:1.1})
gives the Weibull probability distribution function for the dynamic FBM:
\begin{equation} \label{eq:1.3}
P(t;\sigma)=1-exp \left\{ - \nu_0 \left( \frac{\sigma}{\sigma_0}
\right)^\rho t \right\}.
\end{equation}

 The widespread use of Weibull statistics stems from the experimental
fact that real materials follow very closely Weibull probability distribution functions 
for both the strength and the time to failure of the individual elements \cite{daniels}, 
\cite{coleman}, \cite{phoenix}, \cite{oko}. 

Besides the power-law breaking rule, Eq.~(\ref{eq:1.2}), another popular assumption in
composites fracture is the {\it exponential breaking rule},
\begin{equation}  \label{eq:1.4}
\kappa_2(\sigma)=\phi\ exp\left[ \eta \left(
\frac{\sigma}{\sigma_0} \right) \right],
\end{equation}
where $\phi$ and $\eta$ are two positive constants (the amplitude
and the characteristic scale of the exponential function). This
breaking rule has a theoretical support in the apparent necessity
of a Boltzmann factor in the load (stress) for any thermally
activated process (as fracture at the molecular level is
interpreted: \cite{phoenix}). The substitution of
Eq.~(\ref{eq:1.4}) into Eq.~(\ref{eq:1.1}) does not give a Weibull
function. Nevertheless, the established use of the two breaking
rules makes it necessary to take both into account, and so we have
dedicated Section V to the discussion of  the lifetime of sets
under the exponential breaking rule.

        The most critical of the three basic ingredients
of all FBM is the {\it load-transfer rule}, where a great deal of
the physics of the models is hidden. Three end-members are of
interest here: the equal load-sharing (ELS) rule, the local
load-sharing (LLS) rule, and the hierarchical load-sharing (HLS)
rule. In the ELS rule, which can be thought of as a mean-field
approximation, the load supported by failing elements is shared
equally among all surviving elements. In the LLS rule, the load of
failing elements is accommodated by a neighborhood whose exact
definition depends on the geometry and dimensionality of the
underlying lattice. In the HLS rule the scheme of load transfers
follows the branches of a fractal (Cayley) tree with a constant
coordination number. Common to all three load-transfer modalities
is the fact that broken elements carry no load.

        ELS models have been used to predict failure under
tension  in elastic yarns and cables with little or no twist, because in these
arrangements the load supported by a failing fiber or cable strand is shared equally by
all the remaining fibers or strands in the bundle. The conjunction of a loose arrangement
and a load under tension facilitates this global-range load redistribution scheme.

LLS models have their natural field of application in the failure of composite materials, and
more specifically in fiber-reinforced composites with brittle fibers embedded in a stiff
matrix \cite{hull}. There, as fiber breaks appear, the matrix serves the important
function of transferring the shear traction generated in the matrix at the point of a
fiber break to the neighboring fibers, with most of the load going to the nearest
neighbors. This arrangement results in a very short-range load redistribution, both
laterally across fibers and longitudinally along the fiber axis.

        More important from a geophysical point of view and
for this paper is the HLS rule recently introduced by Turcotte and
collaborators in the seismological literature \cite{turcotte}. In
this load-transfer modality the scale invariance of the fracture
process is directly taken into account by means of a hierarchical
load-transfer scheme following the branches of a fractal tree. An
important property of the HLS scheme is that the zone of stress
transfer is equal in size to the zone of failure, and this nicely
simulates the Green's function associated with the elastic
distribution of stress adjacent to a rupture \cite{newman}. The
fractal tree structure used to redistribute loads is a mere
construction useful to envisage the way loads from breaking
elements are transferred to unbroken elements. A basic aspect of
the topology of the hierarchical structure is the number of
elements directly linked together; this defines the coordination,
$c$, of the tree. That is, $c$ fibers could be assembled to form a
bundle which would behave as if it were itself a fiber. Then, $c$
of these second-generation fibers could themselves be assembled to
form a bundle which would act as a third-generation fiber, and so
on. This hierarchical assemblage can be continued indefinitely and
an index $n$ is used to describe the level within the hierarchy
or, equivalently, the height in the tree structure. So $n=0$
refers to the individual elements in the system, $n=1$ refers to
the first level in the tree, etc. For $c=2$, $n=0$ implies
individual elements, $n=1$ implies pairs of elements, $n=2$ pairs
of pairs of elements, etc. Thus, an $n$th order tree with
coordination number $c$ would contain $N=c^n$ elements.

        Although we would stress below the importance of the analytical
solution of the FBM, it is enlightening to show how these models can be solved using 
Monte Carlo techniques because that would facilitate the understanding of parts to 
follow. We will focus on the {\it dynamic} FBM, as this is the type of problem we want to 
address in this paper. Consider a set of $N$ elements arranged on the sites of a lattice. 
The general Monte Carlo recipe goes as follows: (1) Assign random lifetimes $t_i$ to the 
$i=1,..., N$ individual elements, as drawn from Eq.~(\ref{eq:1.1}) under unit load; (2) 
Advance time an amount equal to the lifetime of the shortest-lived non-failed element in 
the set, say $\delta$; (3) Reduce lifetimes of all remaining elements by an amount 
$\kappa(\sigma_i)\delta$, where $\kappa(x)$ is either the power-law or the exponential 
breaking rule; (4) Transfer load from the failing element to other sound elements in the 
set according to a preset load-transfer rule (ELS, LLS, or HLS); (5) Proceed to step 2 if 
at least one element is unbroken, or  end if the system has collapsed; (6) Add together 
all the individual $\delta$s to obtain the time to failure $T$ of that particular 
realization of the system. This way of applying the Monte Carlo method is what we will 
refer to as the {\it standard method}.

        Among the different results that one can obtain from
the analytical or numerical solution of the fiber-bundle models (for a review, see 
\cite{duxbury} ), here we are mainly interested in the asymptotic strength (static FBM) 
and asymptotic time to failure (dynamic FBM) of the system. The asymptotic strength is 
defined as the maximum load that an infinite system can support before all its elements 
break. The asymptotic time to failure or lifetime is the minimum time one has to wait 
until an infinite system collapses by all its elements breaking by fatigue. These are 
themselves important questions from an engineering point of view. Table 1 gathers the 
main asymptotic results for the different FBM, including the dynamic HLS model, to which 
this paper is dedicated. 
It has been known since the work of Daniels \cite{daniels} that
the static ELS fiber-bundle model has a critical point in the
sense that for an infinite system there is a zero probability of
breaking the system when applying a load $\sigma$ less than a
critical value $\sigma_c$, and a probability equal to one to break
the system if the applied load is bigger than the critical load.
This is valid for any probability distribution function satisfying
some very mild conditions \cite{daniels}. The critical strength
$\sigma_c$ quoted in Table 1 is only valid, however, for a Weibull
function. As for the dynamic ELS model, Coleman \cite{coleman}
proved rigorously a comparable result, namely that there exists a
critical time $T_c$ below which an infinite system under dynamic
rules has a zero probability of collapsing and above which the
system collapses with a probability of one. $T_c$ varies with the
assumed probability distribution function, but otherwise the
critical-point result is independent of it (the value quoted in
Table 1 is for a power-law breaking rule). Smith and Phoenix
\cite{smith} give a summary of the main asymptotic results for the
static and dynamic ELS FBM.
 Regarding the asymptotic properties of the LLS
models, the work of Smith and co-workers, and of Harlow, Phoenix
and co-workers has been fundamental. They proved that neither the
static nor the dynamic LLS FBM have a critical point. For the
static case, the strength of the system goes to zero as the size
of the system is increased. More specifically, $\sigma_c \propto
1/lnN$ (conjectured by Harlow, Phoenix and Smith \cite{harlow};
proved by Harlow \cite{harlow2}). For the dynamic case a similar
result holds (conjectured by Tierney \cite{tierney}; proved by Kuo
and Phoenix \cite{kuo}). See \cite{phoenix2} for a review of the
static LLS models.

        The static HLS model was shown to lack a critical point
by Newman and Gabrielov \cite{newman2}. In this case the reduction to zero of strength in 
relation to system size is very slow, $\sigma_c \propto 1/lnlnN $, but strictly speaking 
the strength of an infinite system is zero. The dynamic HLS model was introduced in the 
geophysical literature in reference \cite{newman3}. Afterwards, Newman {\it et al} 
\cite{newman} used this dynamic HLS model with the specific aim of finding out if the 
chain of partial failure events preceding the total failure of the set resemble a 
log-periodic sequence. This was motivated by the amazing fit of this type obtained in 
Ref. \cite{sornette1} to data of the cumulative Benioff strain released in magnitude $>5$ 
earthquakes in the San Francisco Bay area before the October 17, 1989 Loma Prieta 
earthquake. In the analysis of \cite{newman}, it appeared that, contrary to the static 
model, the dynamic HLS model seemed to have a non-zero time to failure as the size of the 
system tends to infinity. This issue was also studied in \cite{sornette2} by using a 
renormalization approach. This behavior seems odd because, as Table 1 shows, there is a 
symmetry, for a specific load-transfer rule, between the static and the dynamic cases: 
the ELS models have a critical point both for the static and the dynamic cases; the LLS 
models have no critical point behavior either for the static or the dynamic case. The 
static HLS model has no critical point, and so it would seem natural that the dynamic HLS 
would not have a critical point either. Here we present an exact iterative method to 
compute the time to failure of sets of elements with a hierarchical modality of 
load-transfer. (A preliminary account of the behavior of the dynamic HLS model under the 
power-law breaking rule has been recently published \cite{gomez}). Due to the fact that 
the exact method is too time consuming to yield useful asymptotic results, we also 
present here rigorous upper and lower bounds to the lifetime of large dynamic HLS sets. 
From the behavior of the lower bound we conclude that the dynamic HLS model has indeed a 
critical point, that is, its time to failure is non-zero for an infinite system. 

This paper is organized as follows. In Section II we review the
continuous formulation of the ELS model given by Coleman
\cite{coleman}. This is useful for understanding the probabilistic
approach used in the rest of the paper. This approach is compared
with the standard method in a Monte Carlo simulation, to
illustrate that both are equivalent. Section III contains the core
of the iterative method to exactly calculate the time to failure
of dynamical HLS sets. The strategy of the juxtaposition of
configurations is explained and the need of defining ``replica''
configurations
 is introduced.
 This leads to the concept of {\it primary diagram} from which the value of the $\delta$s and
 of $T$, for a given $n$, are exactly calculated. From a primary diagram one obtains an
 easier one called {\it reduced diagram}, which is used to build the primary diagram of the
 next level $n+1$.
  In Section IV, aiming at simplification, we introduce the concept of {\it effective diagram},
as the averaged form of a primary. In these diagrams each stage of breaking is
represented by a unique effective configuration whose decay width is obtained by an
appropriate average of the various decay widths existing in the primary. Depending on the
type of means used, upper or lower bounds for the $T$ of the next height are obtained.
    So far the power law breaking rule is used in the quantitative calculations.
Section V is devoted to the exponential breakdown rule specifics.
Two Appendices have been added: In Appendix A we detail the number
of replicas demanded for a general coordination, $c$; in Appendix
B, we show the reason why the three types of means used in the
averaging effectively work to provide rigorous bounds.

\section{THE  ELS MODEL. THE PROBABILISTIC APPROACH} 

In Ref. \cite{coleman}, in the context of his statistical theory
for the time-dependence of mechanical breakdown in bundles of
fibers at constant total load,  Coleman defines an ``ideal bundle"
as one fulfilling precisely the same premises that the ELS model
presented in Section I. Following the work of this author, let us
call $N $ the size of the bundle (or set) at $t=0$ and $n(t)$ the
number of filaments (or elements) which survive to $t$ without
breaking; the lifetime, $T$, of the bundle is defined as the time
required for $n(t)$ to reach zero. We will call $\sigma_o$ the
fixed load attributed to every single element at $t=0$. The
hypothesis of the ELS model implies that the actual load in a
particular unbroken filament at time $t$ is
 \be \label{eq:2.1}
 \sigma = \frac{\sigma_o  N }{n(t) }.
 \ee
 Thus the lifetime of a very large ideal bundle formed by fibers of equal length, $l$, may be
calculated from
 \be  \label{eq:2.2}
  -\frac{dn}{dt}=n\, l\,
\kappa(\sigma) ,
 \ee
where $\kappa(\sigma)$ is a phenomenological function. The product
$l\, \kappa$ is called the hazard rate. In polymeric fibers,
$\kappa$, can be satisfactorily represented by an exponential
function of $\sigma$~:
 \be  \label{eq:2.3}
  \kappa(\sigma) = \frac{ e^{\beta \sigma}}{a}\ ,
 \ee where $a$ and $\beta$ are parameters that
determine the behavior of the fibers under any load $\sigma$.
Eq.~(\ref{eq:2.3}) represents the so called {\it exponential breakdown
rule}. An alternative also widely used is the so called {\it power-law
breakdown rule}:
 \be  \label{eq:2.4}
  k(\sigma)=\frac{1}{a}
\left(\frac{\sigma}{\sigma_o}\right)^\rho \, .
 \ee
From (\ref{eq:2.1}), (\ref{eq:2.2}) and (\ref{eq:2.3}) with
$x=(\beta \, \sigma_o\, N)/n $, and the conditions $n(0)=N,\ n(T)=0$,
one deduces
\be \label{eq:2.5}
T= \frac{a}{l}\ \int_{\beta \sigma_o}^{\infty} \frac{dx\, e^{-x}}{x} = -\frac{a}{l}
E_i(-\beta \sigma_o)\ .
\ee
Using (\ref{eq:2.4}) instead of
(\ref{eq:2.3}) one obtains
\be
\label{eq:2.6} T=\frac{(a/l)}{\rho}\ .
 \ee

Eq.~(\ref{eq:2.2}) is similar to a radioactivity equation in which
$l\cdot \kappa $ stands for the decay rate, $\Gamma$, of one
nucleus. In the ELS case it is not of much interest to lose this
elegant continuous formulation. However for other load-transfer
schemes, such as the HLS, this analogy with radioactivity is
useful, but similar continuous differential equations can not be
formulated anymore. Thus, the discrete version of this
probabilistic philosophy applicable to any load transfer scheme
was developed in Ref.~\cite{gomez98} and will be used throughout
this paper. It represents an alternative to what we have called
the standard method~\cite{newman} commented in Section I, in which
the random thresholds for breaking are assigned at the beginning
and the process of breaking is henceforth completely
deterministic. Both points of view are equivalent. In the
following we will use non-dimensional magnitudes. In particular
\be
\label{eq:2.7} (a/l)=1,\ \sigma_o=1,$ and $\beta\, \sigma_o=1 \ .
\ee Note that this would be equivalent to adopting, with the
notation of Section I, $\nu_o=\sigma_o=\phi=\eta=1$.

In the probabilistic approach~\cite{gomez98}, in each time
increment, defined as
 \be  \label{eq:2.8}
  \delta=\frac{1}{\sum_j \Gamma_j}\ ,
 \ee
 one element of the sample decays. The index $j$ runs along all the surviving
elements.
 Using Eqs. (\ref{eq:2.3}), (\ref{eq:2.4}) and (\ref{eq:2.7}) we have
 \be \label{eq:2.8b}
\Gamma_j=\sigma_j^\rho \ (or\ \Gamma_j=e^{\sigma_j})\ .
 \ee
     The probability of the specific element, $m$, to fail is
\be  \label{eq:2.9}
p_m=\Gamma_m \cdot \delta \ .
\ee
Equation~(\ref{eq:2.8}) is the ordinary link between the
mean time interval for one element to decay in a radioactive sample and the total decay
width of the sample. The time to failure, $T$, of a bundle (set of elements) is the sum
of the $N\ \delta$s.

It is instructive to apply Eq.~(\ref{eq:2.8}) to the ELS case
where $\Gamma_j$ does not depend on $j$ because every surviving
element bears the same load. Here, in the $k$th time step, the  number
of survivors is $n_k=N-k$ and the individual load is $ \sigma_k=
N/(N-k)$. Then for the power law rule
\[
\delta_k=\frac{1}{N-k} \cdot \frac{(N-k)^\rho}{N^\rho} =
\frac{(N-k)^{\rho-1}}{N^\rho} \ ,
\]
with  $k=0,\ ...,\ N-1$ and
\be  \label{eq:2.10}
T=\sum_{k=0}^{N-1} \delta_k
\ee
If $\rho=2,\ T=\frac{1}{2} (1+\frac{1}{N})$. For a general value of $\rho$, using Stolz's
theorem~\cite{stolz} we find
\be  \label{eq:2.11} \lim_{N\rightarrow \infty } T=\frac{(N-1)^{\rho-1}}{N^\rho-(N-1)^\rho}
\longrightarrow \frac{1}{\rho} \ ,
\ee
which coincides with Eq.~(\ref{eq:2.6}).

When dealing with the exponential breakdown rule, proceeding
analogously one easily checks that the sum of the series of
$\delta$s, for sufficiently high $N$, provides the same result as
the exponential integral of Eq.~(\ref{eq:2.5}).

It is also instructive to compare the results obtained from Monte
Carlo simulations in the calculus of $T$ in two ways: (a) by using
the standard procedure, i.e., of assigning random individual
lifetimes at the beginning of each simulation and proceeding
deterministically; or (b) by using a probabilistic point of view,
i.e., from Eq.~(\ref{eq:2.8}) and Eq.~(\ref{eq:2.9}). This
comparison is shown in Fig.~\ref{figure1} for HLS sets of $N=128$
and  $N=512$ elements (with $c=2,\ \rho=2$).  Note the significant
reduction in the dispersion thickness obtained by using this
second method. This contrast tends to decrease for growing $N$ and
growing $\rho$. Obviously, the Monte Carlo strategy can be applied
for any modality of load transfer in the framework of the
probabilistic method. The inconvenience lies in the very essence
of these simulations, i.e., their moderate accuracy and large cost
for large sets. In this paper, we will show how to apply the
probabilistic method to the HLS transfer modality, in order to
obtain an exact algebraic method for the lifetimes, and how to
explore the asymptotic values of $T$ when $N$ tends to infinity.

\section{AN EXACT ITERATIVE METHOD FOR HLS DYNAMICAL  MODELS }

To give a perspective of what is going on in the rupture process
of a hierarchical set we have drawn in Fig.~\ref{figure2} the
three smallest cases for trees of coordination \(c=2\). Denoting
by \(n\) the number of levels, or height of the tree, i.e.
\(N=2^{n}\), we have considered \(n=0,1\) and \(2\). The integers
within parenthesis \((r)\) account for the number of failures
existing in the tree. When there are several non-equivalent
configurations corresponding to a given \(r\), they are labeled as
\((r,s)\), i.e, we add a new index $s$. The total load is
conserved except at the end, when the tree collapses. Referring to
the high symmetry of loaded fractal trees, note that each of the
configurations explicitly drawn in Fig.~\ref{figure2} represents
all those that can be brought to coincidence by the permutation of
two legs joined at an apex, at any level in the height hierarchy.
Hence we call them non-equivalent configurations or merely
configurations. In general, each configuration $(r,s)$ is
characterized by its probability $p(r,s)$, $\sum_{s}p(r,s)
=1$, and its decay width $\Gamma(r,s)$. The time step for
one-element breaking at the stage $r$, is given by
\begin{equation}  \label{eq:3.1}
\delta_{r}=\sum\limits_{s}p(r,s)\frac{1}{\Gamma(r,s)} \label{two} \ .
\end{equation}
This is the necessary generalization of Eq.~(\ref{eq:2.8}) due to
the appearance, for the same $r$, of non-equivalent configurations
 during the decay process of the tree. In cases
of branching, the probability that a configuration chooses a
specific direction is equal to the ratio between the partial decay
width in that direction and the total width of the parent
configuration. And the probability of a given configuration
$p(r,s)$ is given by the sum, extended to all its possible
parents, of the product of the probability of each parent times
the probability of choosing that specific direction.

We will compute at a glance the \(\delta\)s of Fig.~\ref{figure2}
in order to analyze the general case later. To be specific, we
will always use \(\rho=2\). For \(n=0\), we have $\Gamma(0)=1^{2}$
and \(\delta_{0}=\frac{1}{1^{2}}=1=T\). For \(n=1\),
\(\Gamma(0)=1^{2}+1^{2}=2\), $\delta_{0}=\frac{1}{2}$;
\(\Gamma(1)=2^{2}\), $\delta_{1}=\frac{1}{4}$; and hence
\(T=\frac{1}{2}+\frac{1}{4}=\frac{3}{4}\). For \(n=2\),
\(\Gamma(0)=1^{2}+1^{2}+1^{2}+1^{2}=4\),
\(\delta_{0}=\frac{1}{4}\); \(\Gamma(1)=2^{2}+1^{2}+1^{2}=6\),
$\delta_{1}=\frac{1}{6}$. Now we face a branching; the probability
of the transition $(1)\rightarrow(2,1)$ is $\frac{4}{6}$ and the
probability of the transition $(1)\rightarrow(2,2)$ is
$\frac{2}{6}$; on the other hand
$\Gamma(2,1)=2^{2}+2^{2}=8=\Gamma(2,2)$, hence
$\delta_{2}=\frac{4}{6}\cdot\frac{1}{8}+\frac{2}{6}\cdot\frac{1}{8}=\frac{1}{8}$.
Finally $\delta_{3}=\frac{1}{16}$ and the addition of $\delta$s
gives $T=\frac{29}{48}$.

Now we define the {\it replica} of a configuration belonging to a
given \(n\), as the same configuration but with the loads doubled
(this is because we are using \(c=2\)). The replica of a given
configuration will be recognized by a prime sign. In other words
$(r,s)'$ is the replica of $(r,s)$. Note that when a configuration
represents the state of complete collapse, it and its replica are
the same thing. When dealing with the power law breakdown rule,
any decay width, partial or total, related to $(r,s)'$ is
automatically obtained by multiplying the corresponding value of
$(r,s)$ by the common factor $c^{\rho}=2^{\rho}=4$. This also
implies that $p(r,s)=p(r,s)'$. In the exponential rule, this does
not work and the widths of the replicas have to be specifically
calculated (this is explained in Section V). The need to define
the replicas stems from the observation that any configuration
appearing in a stage of breaking $r$ of a given \(n\), is built as
the juxtaposition of two configurations of the level \(n-1\),
including also the replicas of the level $n-1$ as ingredients of
the game. In Fig.~\ref{figure2}, one can observe the explicit
structure of the configurations of $N=4$ (or of $N=2$) as a
juxtaposition of those of $N=2$ (or of $N=1$) and its replicas.
From this perspective, we notice that the total number of
configurations appearing in the fracture process of a tree of
height $n$ (omitting the totally collapsed one), ${\cal N}_n$, is
equal to
 \[   {\cal N}_n= \frac{
{\cal N}_{n-1} \ ( {\cal N}_{n-1} +1) }{2} + {\cal N}_{n-1} \ .
\]
In this formula the first term represents all the possible
combinations (with repetition) of pairs of ordinary configurations
of the height $n-1$. The second term represents the configurations
formed by juxtaposing a collapsed tree of height $n-1$ together
with any of the ${\cal N}_{n-1}$ replicas of the previous height.
Thus, \be  \label{eq:3.3} {\cal N}_n= \frac{ {\cal N}_{n-1} \ (
{\cal N}_{n-1} +3) }{2} \ . \ee Feeding ${\cal N}_0=1$ into
Eq.~(\ref{eq:3.3}), we obtain ${\cal N}_1=2$, ${\cal N}_2=5$,
${\cal N}_3=20$, ${\cal N}_4=230$, ${\cal N}_5=26795$, ${\cal N}_6
\simeq3.59 \times 10^8 $, ${\cal N}_7 \simeq 6.45 \times 10^{16}
$, etc. It is clear that the amount of configurations to deal with
soon constitutes an unsurmountable problem.

    The single-element breaking transitions in configurations of height $n$
can be only of three types. Type {\it a} transitions correspond to
the breaking of one element in a half of the tree while the other
half remains as an unaffected spectator. Type {\it b} transitions
correspond to the decay of the last surviving element in one half
of the tree, which provokes its collapse and the corresponding
doubling of the load borne by the other half. In these two cases,
the transition width coincides with that already obtained when
solving the level $n-1$. Finally, type {\it c} transitions
correspond to the scenario in which one half of the tree has
already collapsed and in the other half one breaking occurs. In
this third case the decay width is that of a replica of the level
$n-1$, which as said before, is a common factor $c^\rho$ times the
ordinary width. This holds for any height $n$ and allows the
computation of all the partial decay widths in a tree of height
$n$ from those obtained in the height $n-1$. This is illustrated
in Fig.~\ref{figure3}, for the three types of transitions for
trees with $n=3$ ($\rho=2$ has been used).

In Fig.~\ref{figure2} and Fig.~\ref{figure3} we have drawn the
different configurations of low $n$ explicitly, that is, by
representing them as small  fractal trees at different stages of
damage. It is convenient, for reasons of economy, to introduce a
symbolic notation for the configurations so that the complete
process of breaking of a tree of height $n$ adopts a more compact
look. This is shown in Fig.~\ref{figure4}. There the different
configurations of $n=3$, are labeled by the integers within the
boxes. The two parentheses at their right, with their respective
integers, represent the two $n=2$ juxtaposed configurations
forming that of level $n=3$. This information of the previous
height will be called the {\it genealogy}. Time is assumed to flow
downwards. The numbers accompanying an arrow connecting two boxes
stand for the decay width of that transition. Coming back to
Fig.~\ref{figure3} one recognizes there that those explicit
transitions are nothing else but what in Fig.~\ref{figure4} is
represented as \fbox{3,1} $\longrightarrow$ \fbox{4,2}, \fbox{3,1}
$\longrightarrow$ \fbox{4,1} and \fbox{4,1} $\longrightarrow$
\fbox{5,1}.  A diagram like that of Fig.~\ref{figure4} is called a
{\it primary}, because it is  formed by the juxtaposition of all
possible configurations of the previous height. Thus, each
configuration belonging to a primary diagram has a specified
genealogy. This allows the computation of all the decay widths of
the diagram. As foreseen in Eq.~(\ref{eq:3.3}), not counting the
totally collapsed configuration, ${\cal N}_3=20$. The sum of all
the partial widths of a parent configuration in a branching is
always equal to the total decay width, $\Gamma$, of the parent.
From this primary width diagram one deduces the probability of any
primary configuration at any stage $r$ of breaking, and
consequently $\delta_{r}$ is obtained using Eq.~(\ref{eq:3.1}).
Finally, by adding all the $\delta$s we calculate $T(n=3)$.

After a primary diagram has been obtained, i.e., after calculating
all its decay widths, it can be simplified. The idea is to fuse,
at each $r$, all the configurations having the same total decay
rate, $\Gamma$. Once fused, these configurations have a
probability equal to the sum of the old probabilities, and
obviously maintain the same $\Gamma$. A primary diagram simplified
in this way will be called a {\it reduced} diagram. An element of
a reduced diagram resulting from a fusion has no genealogy in the
sense that it does not  derive from one but from several
juxtapositions. The genealogy was  used in the calculation of the
primary diagram. The later fusion does not require any other
independent information. To illustrate the concept of what a
reduced diagram is, let us look again at Fig.~\ref{figure2}. For
$n=0$ and $n=1$, for each $r$ there is only one configuration and
hence primary and reduced diagrams are identical. For $n=2$, for
$r=2$ there are two configurations in the primary diagram, but
they have the same width, specifically, for $\rho=2,\ \Gamma
(2,1)=\Gamma (2,2)=8$. Thus these two configurations can be fused
and the resulting effective diagram is a chain of five elements,
i.e., the branching disappears. Performing this task with the
$n=3$ of Fig.~\ref{figure4} one would obtain the reduced diagram
of Fig.~\ref{figure5}. The total number of configurations
appearing in the reduced diagrams, ${\cal N}_n'$, does not derive
from a closed formula as occurs with ${\cal N}_n$. However, it can
easily be derived by means of the computer and obtain: ${\cal
N}_1'=1$,  ${\cal N}_2'=2$, ${\cal N}_3'=10$, ${\cal N}_4'=36$,
${\cal N}_5'=202$, ${\cal N}_6'=1669 $, ${\cal N}_7'=16 408$. The
important point is that one can  use a reduced diagram of the
level $n$ to build a primary of the level $n+1$ obtaining the
exact information of the new level. After calculating that
primary, by fusing again configurations of equal $\Gamma$, one
would obtain the reduced diagram of the height $n+1$.

By iterating this procedure, that is by forming the primary
diagram of the $n+1$ height by juxtaposing the configurations of
the reduced diagram of the height $n$, we can, in principle,
exactly obtain the total time to failure of trees of successively
doubled size. In spite of the great simplification obtained when
using reduced diagrams, the problem of dealing with a vast  amount
of configurations still remains. This fact eventually blocks the
possibility of obtaining exact results for trees high enough as to
be able to gauge the asymptotic behavior of $T$ in HLS sets. A few
examples of exact results, for $c=2,\ \rho=2$   are
$T(n=3)=\frac{63451}{123200}$,
$T(n=4)=\frac{21216889046182831}{46300977698976000}$ $T(n=5)=0.420
823 219 104 814$. The iterative procedure was programed in
Mathematica 3.0 with infinite precision and took 10 minutes CPU
time for $n=5$.

\section{Bounds for the time to failure of the HLS models}

As seen above, whenever in a primary diagram one fuses
configurations of the same $\Gamma$, no information is lost and
the calculation of the time to failure remains exact. In spite of
this simplification the magnitude of the Bayesian problem becomes
huge even when dealing with a moderate $n$.
 That is why we have looked for alternative approximation  procedures to estimate $T$. In fact
the most important goal, as explained in Section I, is to find out
if the $T$ of very large HLS sets tends to zero or, on the
contrary, remains finite. With these points in mind, we have found
that a drastic but appropriate simplification of the primary
diagrams, in which one averages all the configurations of a given
$r$ into a unique configuration with an effective decay width,
leads to obtaining, in the subsequent heights, values of $T$
systematically lower (or higher), than the exact result. As this
fusion leads to only one effective configuration, it will have
probability 1. The value of its decay width will be called $a_r$.
Such ``chain'' diagrams will be called {\it effective diagrams}.
For $n=3$, this is drawn in Fig.~\ref{figure6}.  These effective
diagrams, which substitute the previously defined reduced
diagrams, are used exactly in the same way, i.e., to calculate a
new (approximate) primary diagram of the next height. The economy
obtained by using effective diagrams is obvious. As the number of
effective configurations of a level $n-1$ is $2^{n-1}$, the
primary of height $n$, built from this effective diagram of height
$n-1$, will have a number of configurations ${\cal N}''$, given by
\be
{\cal N}''=\frac{2^{2n-2} + 3\cdot 2^{n-1} }{2}\ ;
\ee
 that is ${\cal N}_1''=2,\ {\cal N}_2''=5,\ {\cal N}_3''=14,\ {\cal
N}_4''=44,\ {\cal N}_5''=152,\  {\cal N}_6''=560 ,\  {\cal
N}_7''=2144 ,$ etc.

It is clear that for $n=0, 1$ and $2$, the reduced diagrams and
the effective diagrams are identical, i.e., $a_r=\Gamma(r)$. The
point is to define $a_r$ for $n\geq 3$ so that the $T(n\geq 4)$
are lower (or higher) than its exact result.

A trivial option is to define \be \label{eq:4.1}
a_r=\Gamma_{max}(r)\  \mbox{ or } \ (\Gamma_{min}(r)) \ee i.e. by
assuming that the only configurations formed during the breaking
of the tree are those of the maximum (minimum) value of $\Gamma$.
As it is easy to foresee, the use of Eq.~(\ref{eq:4.1}) leads to poor
bounds. In fact the lower bound goes quickly to zero. We have
found that good lower bounds are obtained by using effective
diagrams where $a_r$ is the arithmetic mean $(AM)$,
\be
\label{eq:4.2} a_r(AM)= \sum_s p(r,s)\ \Gamma(r,s) \ , \ee or even
better by using the geometric mean $(GM)$,
 \be \label{eq:4.3}
a_r(GM)=\prod_s \Gamma(r,s)^{p(r,s)}\ .
 \ee
  Good higher bounds are obtained from the harmonic mean $(HM)$, \be
\label{eq:4.4} a_r(HM)= \frac{1}{ \sum_s p(r,s)\,
\frac{1}{\Gamma(r,s)} } \ .\ee

In Appendix B, we analyze why bounds result. Note that given the primary diagram of a 
height $n$, which leads to $c^n\ \delta$s, the elements forming the effective diagram 
defined with the aim of obtaining higher bounds are exactly $a_r=1/\delta_r$. The fact 
that the $a_r(i),\ i=AM,GM,$ and $HM$   lead to bounds in the form explained above is in 
qualitative concordance with the inequality \be \label{eq:4.5} \Gamma_{min}(r) \leq 
a_r(HM) \leq a_r(GM) \leq a_r(AM) \leq \Gamma_{max} (r) \ , \ee which is always a 
mathematical fact. The bounds obtained from these formulae for $c=2, \rho=2$ are plotted 
in Fig.~\ref{figure7}, together with points representing Monte Carlo results. As the 
bounds based on the $GM$ and on the $HM$  are the most stringent, they will be called 
$T_l$ and $T_h$ respectively. The detailed behavior of $T_l$ has been analyzed in a 
log-normal plot of \(T_l-T_{l,\infty}\) against the number \(n\) of levels of the tree. 
\(T_{l,\infty}\) is a constant obtained from a fit of the data points to the exponential 
function \(a e^{-b(n-n_{0})}\) shifted downwards by this amount \(T_{l,\infty}\) (\(a\), 
\(b\) and \(n_{0}\) are three fitting parameters of no interest here). We have performed 
a careful sensitivity analysis of the four-parameter exponential fitting because the 
success of this exponential decay to a non-zero limit is the hallmark of the claim. 
Table\ \ref{table1} records \(T_{l,\infty}\) obtained from an exponential fit to the last 
\(k\) data points. The first \(T_{l,\infty}\) column is for a fit using up to a maximum 
level of \(n=20\) (hence the notation \(n_{max}\) in the table); the second 
\(T_{l,\infty}\) column is for the same fit but dropping the \(n=20\) value; for the 
third \(T_{l,\infty}\) column we have also dropped the \(n=19\) data point. It is clear 
from the trend in the three \(T_{l,\infty}\) columns that a saturation towards 
\(T_{l,\infty}=0.32537\ \pm 0.00001\) occurs when using only information of big trees to 
perform the non-linear fitting. A similar analysis of $T_h$, leads to 
$T_{h,\infty}=0.33984\ \pm 0.00001$. The quality of this exponential fit is also shown in 
Fig.~\ref{figure8}. Similar fittings of the Monte Carlo data points are inconclusive, due 
to the intrinsic noisiness of the MC results and the limited size of the simulated sets 
($N<2^{16}$ elements). What this result implies is that a system with a hierarchical 
scheme of load transfer and a power-law breaking rule $(c=2,\rho=2)$ has a time to 
failure for sets of infinite size, $T_{\infty}$, such that $0.32537 \leq T_{\infty} \leq 
0.33984$. Thus, there is an associated zero-probability of failing for $T<T_{\infty}$ and 
a probability equal to one of failing for \(T>T_{\infty}\). The critical point behavior 
is thus numerically confirmed. 

\section{Exponential breaking rule}

When dealing with the exponential breaking rule in the probabilistic approach, one has to
use Eq. (\ref{eq:2.3}) for the hazard  rate function. For the specific value  of the
parameters as fixed in Eq. (\ref{eq:2.7}), we have
 \be \label{eq:5.1}
 \Gamma_j=e^{\sigma_j} \ .
\ee
 The problem with this breaking rule is that the values of the decay widths appearing in a
diagram where the loads are doubled, i.e., in a diagram replica,
are not obtained by multiplying the normal ones by a fixed
constant, as occurred with the power-law breaking rule. This is
easily checked in Fig. \ref{figure9} where the primary  diagram
for $n=2$ and its replica are shown. The notation is equal to that
of Section II. The values of the decay widths here are dictated by
Eq. (\ref{eq:5.1}). Several comments are in order. We see that the
structure of the diagrams is equal to those appearing for the
power law case because this is independent of the breaking rule
assumed.  In Fig. \ref{figure9} (b) we have drawn explicitly a
replica, that is a diagram in which the system instead of starting
with individual loads $\sigma_o =1$, starts with doubled
individual loads $\sigma_o'=2 \sigma_o =2$. We also see that, just
in the same way as the quantitative calculation of the primary of
Fig. \ref{figure9} (a) required the knowledge of the information
of the previous height and of its replicas, the calculation of the
diagram of Fig. \ref{figure9} (b) requires the knowledge of the
primed elements and the double-primed elements (i.e. with loads
multiplied by $4$) of the previous height. Thus, suppose that we
want to calculate $T$ up to the height $n=4$. This demands the
knowledge of the reduced diagram of $n=3$ and of its replica. We
will denote them by $\{3\}$ and $\{3\}'$. To obtain $\{3\}$ we
need to know $\{2\}$ and $\{2\}'$, and to obtain $\{3\}'$ we need
to know $\{2\}'$ and $\{2\}''$. Going backwards up to $n=0$, we
observe that the scheme of information needed looks like the
following triangular array :
 \[
\begin{array}{lllll}
\{0\}=e & \{0\}'=e^2  & \{0\}''=e^4  & \{0\}''' = e^8 & \{0\}''''=e^{16} \\ \{1\} &
\{1\}' & \{1\}'' & \{1\}'''  &    \\ \{2\} & \{2\}'  & \{2\}''  &      &    \\ \{3\} &
\{3\}' & & &
\\ \{4\} & &   &   &
\end{array}
\]

In other words, to obtain the $T$ of a given height $n$ we have to
explicitly calculate the primary diagrams of the previous values
of $n$, starting from $n=0$, up to a loading $n$ times the usual
diagram with $\sigma_o=1$. The primary diagrams of low $n$ are
calculated at once, hence the extra work with respect to the power
law case is not too much. When explaining the above mentioned
triangular array we were referring to exact primary diagrams,
taking for granted that our aim was to obtain exact results.
Obviously this  changes if our aim is the calculation of bounds;
then one would proceed by averaging widths for each $r$, that is
by calculating means and dealing with effective diagrams.

In Fig. \ref{figure10}, we have drawn the lifetimes, $T$, for trees of height $n$. The 
circles are results obtained from Monte Carlo simulations. For the exponential breaking 
rule, the lower bound based on the arithmetic mean, curve (3) goes to zero. Thus the only 
lower bound that remains useful is that based on the geometric mean. Again, it will be 
called $T_l$. By fitting the data $T_l$ by an exponential function of the form $T_l=T_{l, 
\infty} + a e^{-b(n-n_o)} $ we observe a clean saturation of the asymptotic time to 
failure towards $T_{l, \infty}=0.05285 \mp 0.00001$; analogously, we obtain $T_{h, 
\infty}=0.08825 \mp 0.00001$. Hence, the critical point behaviour is also numerically 
confirmed for the exponential breaking rule. 

\section{ CONCLUSIONS }

In this paper the time to failure, $T$, of hierarchical
load-transfer models of fracture has been studied. Initially we
have explained in detail the so called probabilistic approach to
load-transfer dynamical models as opposed to the standard
approach, in which random lifetimes are assigned to the elements
of the set and the process of fracture evolves deterministically.
We have emphasized that when viewed from the probabilistic point
of view, the calculation of $T$ is analogous to the computation of
the total decay time of a radioactive sample. In fact the
terminology of radioactivity appears throughout this paper. We
have shown that the calculation of $T$, using Monte Carlo
simulations, has a smaller dispersion if one adopts the
probabilistic approach.

Then, we have devised an exact method to compute $T$ of
hierarchical structures of size $N=c^n$. The number of elements of
the set is $N$, $c$ is the coordination of the tree and $n$ the
height of the fractal tree. The method is iterative, i.e., for a
given $c$, it allows the computation of a tree of height $n+1$
once one has calculated a tree of height $n$. In this context, the
sentence ''a tree is calculated" means that one knows the value of
all the partial decay widths between all possible configurations
appearing during the breaking process of that tree. Once this
information is known, one easily calculates the probability of
reaching each configuration, and the individual values of each
$\delta$, i.e., the one-element breaking time. The key of the
method derives from the observation that the structure of the
configurations of the $n+1$ type are a mere juxtaposition of $c$
configurations of the $n$ type. In this juxtaposition, the so
called replicas also play a role. The quantitative information of
how replica configurations behave is explained for the two
relevant breaking rules: the power law and the exponential. In the
power law  breaking rule any decay width of a replica is just a
common factor times the original value. In the exponential
breaking rule, on the contrary, the decay widths of  replicas have
to be individually calculated.

The iterative process, including the information of the replicas,
can be easily processed by a computer. It allows the exact
calculation of $T$ for moderate heights $n$. An exact
simplification, denoted as reduction, is introduced to diminish
the magnitude of the information to deal with. But even with the
reduction trick, it is difficult to surpass, say $n=7$ for $c=2$.
Higher values of the coordination implies smaller values for the
accessible height.

Thus we conclude that exploring the behavior of $T$, for large $n$, dealing with exact 
results, is impossible. For this reason we have turned our interest towards developing 
simple approximate methods which can, however, provide interesting information on the 
asymptotic value of $T$. In this context appears the idea of obtaining bounds for $T$. It 
is found that by performing adequate averages of the decay widths appearing at each stage 
of breaking of a height $n$, the value of $T$ obtained in the next height $n+1$ is 
systematically lower (or higher) than what the exact result would be. In one Appendix we 
have given details of why bounds result. As the results obtained from the bounds reach 
values beyond $n=17\ (c=2)$, one is able to explore their asymptotic behavior by a 
careful exponential fitting which provides clear numerical evidence (although non 
rigorous) that for $c=2$, $T$ tends to a non-zero value when $n$ tends to infinity. This 
conclusion is obtained for both the power-law and the exponential breaking rules. For the 
power law hazard rate, a proof was given in \cite{newman4}. Invoking conventional 
universality- class arguments one deduces that this non-zero limit holds for hierarchical 
structures of any coordination. 

\acknowledgments

A.F.P is grateful to J. As\'{\i}n, J. Bastero, J.M. Carnicer and L. Moral for clarifying 
discussions. M.V-P. thanks Miren Alvarez for discussions. Y.M thanks the AECI for 
financial support. This work was supported in part by the Spanish DGICYT. 

\appendix

\section{ON THE REPLICAS}
 We have seen in Section III that the knowledge of the reduced configuration of a
level $n-1$ is not enough to obtain the primary diagram of the
level $n$; we also have to know the replica of the $n-1$ level.
Expressed in the singular, this sentence is  misleading. In fact
it only holds for $c=2$. One can easily check that for a general
$c$ , the number of replicas required, $m$, is
 \be \label{eq:A1}
 1\leq m \leq (c-1) \ .
 \ee
In the case of the power-law breaking rule, any decay width of these replicas would be
obtained by multiplying its normal value by the factor
 \be \label{eq:A2}
 \left( \frac{c}{c-m}
\right)^{\rho} \ee while as seen in Section V, the exponential
breaking rule demands the individualized calculations of each
replica, with its corresponding extra loading. As an example
beyond the usual $c=2$, let us consider for the case $c=4$ the
process of breaking up to the collapse of the two minimum trees
$n=0$ and $n=1$. Using a self-explanatory notation, we
have:\\[2mm] $n=0$ \ \ \fbox{1} $\longrightarrow$ \fbox{0}\\[2mm]
$n=1$ \

\parbox{6ex}{
   \fbox{1} \fbox{1} \\
   \fbox{1} \fbox{1}  }
$\longrightarrow$
\parbox{7ex}{
   \raisebox{-1mm}[0mm][-1mm]{
\hspace{-2mm} \fbox{\rule[0mm]{0cm}{0.8em}0}} \fbox{\tiny $\frac{3}{4}$} \\
   \ \fbox{\tiny $\frac{3}{4}$}    \fbox{\tiny $\frac{3}{4}$}   }
$\longrightarrow$
\parbox{6ex}{
   \fbox{0} \fbox{0} \\
   \fbox{2} \fbox{2}  }
$\longrightarrow$
\parbox{6ex}{
   \fbox{0} \fbox{0} \\
   \fbox{0} \fbox{4}  }
$\longrightarrow$
\parbox{6ex}{
   \fbox{0} \fbox{0} \\
   \fbox{0} \fbox{0}  }\\[2mm]
We see that the solution of the height $n=1$, demands the
information of the decay width of \fbox{1} but also that of
\raisebox{0.8mm}[0mm][0mm]{ \fbox{\tiny $\frac{3}{4}$}}, of
\fbox{2} and of \fbox{4}; i.e. for $c=4$ the iterative method
requires the knowledge of three replicas, as foreseen in
Eq.~(\ref{eq:A1})

\section{ON WHY BOUNDS RESULT}

Following the arguments of section III and IV, one easily sees that the first $\delta$ in
which there must be a discrepancy between the exact result and the approximate results
coming from the use of effective diagrams is the $\delta_3$ of $n=4$.
 For $c=2,\ \rho =2$, we obtain
\begin{eqnarray} \label{eq:B1}
 \delta_3(HM)&=&\frac{128}{2893}=0.0442447 \nn \\
\delta_3(exact)&=&\frac{17}{385}=0.0441558  \nn \\
\delta_3(GM)&=&\frac{1}{3} \left[
\frac{1}{11}+\frac{1}{8+20^{2/5}\cdot 14^{3/5}} \right]=0.0441074
\\ \delta_3(AM)&=&\frac{59}{1342}=0.0439642 \nn
\end{eqnarray}
 This results from the fusion of
the two configurations \fbox{3.1} and \fbox{3.2} of Fig.
\ref{figure5}, which have a different $\Gamma$. To clarify why
bounds result, let us analyze this point from a general
perspective. In Fig. \ref{figure11} is drawn the top of a reduced
diagram of height, say, $n$, and at its right the corresponding
effective diagram. We assume $\Gamma_1\neq \Gamma_2$,
$a_2=\gamma_1+\gamma_2$, and $i=AM,\ GM$ or $HM$,
 \bea \label{eq:B2}
 a_{3,\ AM}&=&\frac{\gamma_1}{a_2} \Gamma_1 +\frac{\gamma_2}{a_2} \Gamma_2 \ ,\nn \\
 a_{3,\ GM}&=& \Gamma_1 ^{\frac{\gamma_1}{a_2}}  \cdot \Gamma_2 ^{\frac{\gamma_2}{a_2}} \ ,   \\
a_{3,\ AM}&=& \frac{1}{ \left( \frac{\gamma_1}{a_2} \right) \frac{1}{\Gamma_1} + \left(
\frac{\gamma_2}{a_2} \right) \frac{1}{\Gamma_2} } \ .\nn
 \eea

From the reduced diagram we obtain the top of the corresponding
primary of the height $n+1$. This is shown in Fig. \ref{figure12},
and from the effective diagram one obtains the top of the primary
shown in Fig. \ref{figure13}. Now let us compute the exact
$\delta_3$ coming from Fig. \ref{figure12} to be compared with
that (approximate) coming from Fig. \ref{figure13}. In Fig.
\ref{figure12}, we have
 \begin{eqnarray*}
p(3.1)&=&\frac{a_1}{a_0+a_1} \cdot  \frac{\gamma_1}{a_0+a_2} \ , \\
p(3.2)&=&\frac{a_1}{a_0+a_1} \cdot  \frac{\gamma_2}{a_2+a_2} \ ,\\
p(3.3)&=&\frac{a_0}{a_0+a_1} +\frac{a_1}{a_0+a_1} \cdot \frac{a_0}{a_2+a_2} \ .
\end{eqnarray*}
 Thus
\be \label{eq:B3}
 \delta_3(\mbox{exact})= p(3.1) \frac{1}{a_0+\Gamma_1} + p(3.2)
\frac{1}{a_0+\Gamma_2} + p(3.3) \frac{1}{a_1+a_2} \ .
 \ee
  Analogously, in Fig. \ref{figure13},
\begin{eqnarray*}
 p(3.1)&=&\frac{a_1}{a_0+a_1} \cdot  \frac{a_2}{a_0+a_2}  \ ,\nn \\
  p(3.2)&=&\frac{a_0}{a_0+a_1}
   +\frac{a_1}{a_0+a_1} \cdot  \frac{a_0}{a_2+a_2} \ ,
\end{eqnarray*} and
 \be \label{eq:B4}
 \delta_{3,i} = p(3.1) \frac{1}{a_{3,i}+a_0} +
p(3.2) \frac{1}{a_1+a_2}\ .
 \ee
 As the third term of Eq. (\ref{eq:B3}) coincides with the
second of Eq. (\ref{eq:B4}) let us reorder Eq. (\ref{eq:B3}),  giving
 \[
\delta_3(\mbox{exact})- p(3.3) \frac{1}{a_1+a_2}=\frac{a_1}{(a_0+a_1)(a_0+a_2)}
\left(\frac{\gamma_1}{a_0+\Gamma_1}+\frac{\gamma_2}{a_0+\Gamma_2} \right)
 \]
  and similarly in Eq.~(\ref{eq:B4}), obtaining
  \[ \delta_{3,i}- p(2.2)
\frac{1}{a_1+a_2}=\frac{a_1\cdot a_2}{(a_0+a_1)(a_0+a_2)} \left(\frac{1}{a_{3,i}+a_0}
\right) \ .
 \]
 To simplify the comparison, let us define two new functions, \\
 \bea 
\Delta_3(\mbox{exact})&\equiv& \left( \delta_3(\mbox{exact})-
p(3.3) \frac{1}{a_1+a_2} \right)  \frac{(a_0+1)(a_0+a_2)}{a_1}
=\frac{\gamma_1}{a_0+\Gamma_1}+\frac{\gamma_2}{a_0+\Gamma_2} \ ,
\label{eq:B5} \\ \Delta_3(i)&\equiv& \left( \delta_{3,i}- p(2.2)
\frac{1}{a_1+a_2} \right) \frac{(a_0+1)(a_0+a_2)}{a_1}
=\frac{a_2}{a_0+a_{3,i}} \ . \label{eq:B6}
 \eea
  Thus, the $\Delta$s represent
the $\delta$s after adding an equal term, and multiplied by an
equal factor. It is interesting to observe the effect produced by
the fusion of \fbox{3.1} and \fbox{3.2} from an algebraic point of
view: the sum of the two fractions of Eq. (\ref{eq:B5}) has
converted into the fraction at the right of Eq. (\ref{eq:B6}). In
the case of $\Gamma_1=\Gamma_2$, $\Delta_3$(exact)$=\Delta_3(i)$.
In other words, if $\Gamma_1=\Gamma_2$ that fusion is exact.

To deal with the Arithmetic Mean, let us define
 \be \label{eq:B7}
f(x)=\frac{a_2}{a_o+x} \ , \ee $(a_o,\ a_2 > 0)$ which  is
concave; this  implies that $ f(\lambda_1 \Gamma_1+ \lambda_2
\Gamma_2) \leq \lambda_1\, f(\Gamma_1)+ \lambda_2\, f(\Gamma_2)$
where $\lambda_1 \equiv \frac{\gamma_1}{a_2},\ \lambda_2 \equiv
\frac{\gamma_2}{a_2},\ \lambda_1 +\lambda_2=1$ and therefore
\[
\frac{a_2}{ a_0+\left[ \frac{\gamma_1}{a_2}\frac{1}{\Gamma_1} +
\frac{\gamma_2}{a_2} \frac{1}{\Gamma_2} \right]} \leq
\frac{\gamma_1}{a_0+\Gamma_1} + \frac{\gamma_2}{a_0+\Gamma_2} \ ,
\]
which means that $\Delta_3 (AM)\leq \Delta_3 (\mbox{exact})$ and
therefore $\delta_3(AM) \leq \delta_3 (\mbox{exact})$.

To deal with the Geometric Mean let us in Eq.~(\ref{eq:B7})  make
the change of variable $z=\log x$, then
\[
f(x)=\frac{a_2}{a_2+x}\equiv g(z)=\frac{a_2}{a_0+ e^z} \ ,
\]
which is also a concave function in $z$. Hence we have
\[
\lambda_1 \frac{a_2}{a_2+ e^{z_1}}+ \lambda_2 \frac{a_2}{a_2+
e^{z_2}} \geq \frac{a_2}{a_0+ e^{(p_1 z_1+p_2 z_2)} }\ ;
\]
\[
\frac{\gamma_1}{a_0+\Gamma_1} + \frac{\gamma_2}{a_2+\Gamma_2}  \geq
\frac{a_2}{a_0+ (e^{\log z_1})^{\lambda_1}\, (e^{\log z_2})^{\lambda_2}} =
\frac{a_2}{a_0+\Gamma_1^{\ p_1} \Gamma_2^{\ p_2}} \ .
\]
Thus  $\Delta_3 (GM)\leq \Delta_3 (\mbox{exact})$ and $\delta_3(GM) \leq \delta_3(\mbox{exact})$.

Finally, for the higher bound we will make in Eq. (\ref{eq:B7})
the change of variable $z=\frac{1}{x}$ :
\[
f(x)=\frac{a_2}{a_2+x}\equiv h(z)=\frac{a_2 z}{a_0 z+ 1} \ .
\]
$h(z)$ is a convex function in $z$, therefore
\[
h(\lambda_1 z_1+ \lambda_2 z_2) \geq \lambda_1\, h(z_1)+ \lambda_2\, h(z_2) \ ,
\]
and in terms of our ordinary variables this implies
\[
\Delta_3 (H.M.) \equiv f \left( \frac{1}{\frac{p_1}{\Gamma_1}+ \frac{p_2}{\Gamma_2} }\right) \geq
p_1\, f(\Gamma_1)+ p_2\, f(\Gamma_2) =\Delta_3 (\mbox{exact})
\]
and hence $\delta_3(H.M.) \geq \delta_3 (\mbox{exact})$.

Note that the argument presented is valid for the $\delta_3$ of
any $n$ of coordination $c=2$, and for both the power-law breaking
rule and for the exponential breaking rule.


\newpage
\begin{table}
\caption{Main asymptotic results for the three standard modalities
of FBM in the static and dynamic cases.}
\begin{tabular}{lccc}
 & ELS & LLS & HLS \\ \hline
       & CRITICAL POINT         & NO CRITICAL POINT          & NO CRITICAL POINT \\
Static & $\sigma_c=e^{-1/\rho}$ & $\sigma_c \propto 1/lnN$ &
$\sigma_c \propto 1/lnlnN$ \\
       & Daniels (1945)         & Harlow (1985)              & Newman and Gabrielov (1991) \\
       & CRITICAL POINT         & NO CRITICAL POINT          & {\bf CRITICAL POINT}    \\
Dynamic& $T_c=1/\rho$           &                            &
\\
       & Coleman (1958)         & Kuo and Phoenix (1987)     &  This paper
\end{tabular}
\label{table1}
\end{table}

\begin{table}
\caption{Sensitivity analysis of the exponential decay fitting to
the lower bound results.}
\begin{tabular}{dddddd}
\multicolumn{2}{c}{$n_{max}=20$}&\multicolumn{2}{c}{$n_{max}=19$}&\multicolumn{2}{c}{$n_{max}=18$
}\\ k&$T_{l,\infty}$&k&$T_{l,\infty}$&k&$T_{l,\infty}$\\
\tableline 19&0.32575&18&0.32579&17&0.32585\\
18&0.32551&17&0.32553&16&0.32555\\
17&0.32541&16&0.32542&15&0.32542\\
16&0.32537&15&0.32537&14&0.32538\\
15&0.32536&14&0.32536&13&0.32536\\
14&0.32536&13&0.32536&12&0.32536\\
13&0.32536&12&0.32536&11&0.32536\\
12&0.32537&11&0.32537&10&0.32536\\
11&0.32537&10&0.32537&9&0.32537\\ 10&0.32537&9&0.32537&8&0.32537\\
9&0.32537&8&0.32537&7&0.32537\\ 8&0.32537&7&0.32537&6&0.32537\\
7&0.32537&6&0.32537&5&0.32537\\ 6&0.32537&5&0.32537\\ 5&0.32537\\
\end{tabular}
\label{table2}
\end{table}

\newpage

\begin{figure} \caption{Comparison of Monte Carlo simulations; the broad distributions 
come from using the standard approach, and the thinner distributions come from using the 
probabilistic approach. The time to failure is plotted in dimensionless units.} 
\label{figure1} \end{figure} 

\begin{figure}
\caption{Breaking process for the three smallest trees of
coordination $c=2\ (N=1,\ 2,\ 4)$. The integers in parenthesis
($r$) represent the number of breakings occurred. The $\delta$s
stand for the time elapsed between successive individual breakings
and the numbers under the legs indicate the load they
bear.}\label{figure2}
\end{figure}

\begin{figure}
\caption{Calculation of three partial decay widths in $n=3$, from
the information obtained in $n=2$.} \label{figure3} \end{figure}

\begin{figure} \caption{Symbolic representation of the gradual rupture of the tree of height
$n=3$ ($c=2,\ \rho=2$). Time flows downwards.} \label{figure4} \end{figure}

\begin{figure}
\caption{Result of ``reducing" the diagram of Fig.~\ref{figure4}.}\label{figure5} \end{figure}

\begin{figure}
\caption{``Effective'' diagram for $n=3$.} \label{figure6} \end{figure}

\begin{figure} \caption{Dimensionless lifetime, $T$, for a fractal tree of height $n$. 
The small circles are obtained from Monte Carlo simulations. Lines 4 and 1 are higher 
bounds based on $\Gamma_{min}$ and the HM respectively. Lines 2, 3 and 5 are lower bounds 
based on the GM, AM and $\Gamma_{max}$ respectively.} \label{figure7} \end{figure} 

\begin{figure}
\caption{Visualization of the exponential fittings to the results
obtained by using Geometric and Harmonic means.} \label{figure8}
\end{figure}

\begin{figure} \caption{Primary diagrams for the exponential breaking rule ($n=2$).}
\label{figure9} \end{figure}

\begin{figure} \caption{Results for $T$ from trees of height $n$, with the exponential breaking
rule. The small circles correspond to Monte Carlo simulations.
Lines 2 and  3 correspond to lower bounds based on GM and AM
respectively.} \label{figure10}
\end{figure}

\begin{figure} \caption{Top of a failure diagram down to the fourth decay stage. (a) represents
the reduced diagram and (b) the corresponding effective diagram.} \label{figure11} \end{figure}

\begin{figure} \caption{Top of the primary diagram built from the reduced diagram drawn in
Fig.~\ref{figure11} (a).} \label{figure12} \end{figure}

\begin{figure} \caption{Top of the primary diagram built from the effective diagram drawn in
Fig.~\ref{figure11} (b).} \label{figure13} \end{figure}


\end{document}